\documentclass{arXiv}
\usepackage{natbib}

\def\la{\langle}
\def\ra{\rangle}
\def\be{\begin{equation}}
\def\ee{\end{equation}}
\def\bea{\begin{eqnarray}}
\def\eea{\end{eqnarray}}

\copyrightyear{2008}
\pubyear{2008}

\begin{document}
\firstpage{1}

%\title[Knowledge-Integrated SAP/PTM Capable Peptide Identification]
\title[Knowledge-Integrated Peptide Identification]
{RAId\_DbS: A Mass-Spectrometry Based Peptide Identification Web Server with Knowledge Integration}
%{Knowledge Integrated, SAP Capable and PTM Allowed 
% Peptide Identification Web Server based on RAId\_DbS}
\author[Alves {\it et. al.}]{Gelio Alves, 
 Aleksey Ogurtsov, and Yi-Kuo Yu\footnote{to whom correspondence should be addressed: yyu@ncbi.nlm.nih.gov}}
\address{National Center for Biotechnology Information, 
National Library of Medicine, NIH, Bethesda, MD  20894  \vspace{-0.3in}
}
\maketitle

\begin{abstract}

\section{Summary:} 
In anticipation of the individualized proteomics era and the need
 to integrate knowledge from disease studies, we have 
 augmented our peptide identification software RAId\_DbS to take into account
 %while analyzing a tandem mass spectrum,
annotated single amino acid polymorphisms, post-translational modifications, 
  and their documented disease associations while analyzing a tandem mass spectrum. 
% Although it is possible to include 
%  other species, the database constructed 
% contains only 
% This feature is currently limited to human proteins. 
%Through our database construction,
% only a mimimal data expansion is needed to accommodate polymorphisms as well
% as post-translational modifications. The same approach may be used to 
% alleviate the generic data size expansion problem associated with  
% considerations of isoforms, polymorphisms, and post-translational modifications.  
% Not only have we incorporated all the single amino acid polymorphisms 
% and post-translational modifications documented in 
% SwissProt database, 
To facilitate new discoveries, RAId{\_}DbS allows users
 to conduct searches permitting {\it novel} polymorphisms.
%When the underlying peptide is actually
% a polymorphous variant of a peptide in the standard protein database, 
% RAId\_DbS can now identify the variant peptide correctly instead of reporting no significant hits.  
 \section{Availability:}
The webserver link is http://www.ncbi.nlm.nih.gov/ /CBBResearch/qmbp/raid{\_}dbs/index.html.
The relevant databases and  binaries of RAId\_DbS for Linux, Windows, and Mac OS X are 
available from the same web page. 

\section{Contact:} \href{yyu@ncbi.nlm.nih.gov}{yyu@ncbi.nlm.nih.gov}
\end{abstract}

\subsection*{Introduction}
Like single nucleotide polymorphisms (SNPs) that occur roughly every $300$ base pairs~\citep{Collins_98}, 
single amino acid polymorphisms (SAPs) also differentiate individuals from one another. 
In addition to from nonsynonymous SNPs, SAPs may result from post-transcriptional regulations such as 
 mRNA editing. SAPs together with post-translational modifications (PTMs)
 often distinguish healthy/diseased forms of proteins. Integration of
 this annotated, disease-related knowledge with data analysis facilitates speedy, dynamic information retrieval 
 that may significantly benefit clinical laboratory studies. 
% Marching towards an era of individualized
% proteomics, proteomics analysis tools are expected to take into account knowlege information from different sources.   

To incorporate knowlege information within peptide searches, we start by constructing a human 
 protein database where information about annotated SNPs, SAPs, PTMs, and their
 disease associations (if any) are integrated.  
We have also modified our peptide identification 
 software RAId{\_}DbS~\citep{RAId_DbS} to take into account this additional information while
 performing peptide searches. %As will be described in the next sections,
% the database constructed together with the modifications implemented in 
% RAId{\_}DbS minimize the generic problem associated with incorporations 
%of isoforms, SAPs, and PTMs in database searches. 
%This should be contrasted with the previous work
 %of \citet{Mann_2007} where only SAPs are included without integration of disease information.
 Consequently, part of our work may be considered an improvement over that
 of \citet{Mann_2007} who extended the human protein database to include
  only SAPs but without PTMs and without integration of disease information.

Besides using our web server, a user may also download a standalone 
 executable to be installed on her/his local machines. Once a user chooses to do so,
 she/he will find an important feature of the standalone version: the flexibility for
 users to add their own SAP and/or PTM information to various proteins
 they are interested in and even to add new protein sequences to the database.

\subsection*{Implementation Summary}
In addition to giving a brief introduction to our software RAId\_DbS and its augmentation,  
 we focus in this section on explaining how we accommodate the SAPs, PTMs, and their disease associations in our database. Appropriate comparison to existing approaches will also be discussed. 
% Some relevant details will be provided in the supplementary information.  
Prior to database construction, we perform a {\it information-preserved} protein clustering (see supplementary information).   
\vspace{-0.08in}
\subsubsection*{Database Construction}
To minimize inclusion of less confident annotations, we only keep the SAPs and PTMs that are
 consistently documented in more than one source. 
 For example, for proteins with Swiss-Prot accession number,
 we only keep the SAPs and PTMs that are annotated both by Swiss-Prot and GeneBank. 
 For proteins without Swiss-Prot accession numbers, the retentions of SAPs and
 PTMs are described in the supplementary information.

 A typical sequence in our augmented human protein database carries with it
 annotated SAPs and PTMs in a simple format, see Figure~1 and its caption.
 Our data format minimizes redundancy.  
 For example, if a single site contains two SAPs, construction method proposed 
 by \citet{Mann_2007} will demand two almost identical partial sequences, each may 
  be several tens of amino acids in length, be appended after the primary sequence, 
  while in our case it only takes up a few additional bytes.
  The compactness of our database becomes obvious when incorporating the information
  of two nearby sites, each containing several 
  annotated SAPs and PTMs, into the database. In our construction, we only need
  a few additional bytes. But in other approaches, it may introduce a 
  combinatorial expansion due to including/excluding and pairing of different 
  variations at both sites along with the flanking peptides. 
 Another key difference between our method and other database methods is that
 we do not need to limit the number of enzymatic miscleavages. 
 
 When needed,
 users of RAId{\_}DbS may modify the database, add new sequences, 
 or even create their own databases following the same 
 format. There is a separate information file that contains the protein accession numbers, 
 detailed SAP and PTM information, and disease associations. If one wishes to add additional 
 SAPs or PTMs, one simply updates both the ASCII database file as well as the information file.  
When reporting a hit with annotated SAPs or PTMs, RAId{\_}DbS  automatically reports
the corresponding detailed information and disease association if it exists.

\begin{table*}[th!]
\processtable{Example search results of augmented RAId{\_}DbS. \label{tab.1} \vspace{-0.1in}}
{\scriptsize
\begin{tabular}{rllllll}\toprule
(a) $E$-value & $P$-value & Peptide & Mol. Wt. & Protein ID & Novel SAP & Disease \\
1.184e-01& 1.744e-05 & RTKLKDC\ldots KIAR & 2897.500 &(NP{\_}114412;\ldots ;Q9H2L5) & disabled & \\
%7.116e-01&1.628e-03&KANSMEL\ldots ELAR&2899.450&(NP{\_}000489;\ldots;P19099) & & \\
\vdots & \vdots & \vdots & \vdots & \vdots & \vdots & \vdots  \\
4.084e+00 & 9.345e-03& KQQELAA\ldots VSSR & 2898.520 & (NP{\_}072096;\ldots ;O75420) & disabled & \\
\botrule  \toprule
(b) $E$-value & $P$-value & Peptide & Mol. Wt. & Protein ID & Novel SAP & Disease \\
3.977e-07 & 1.834e-10 & KsVEEYANCHLAR & 1448.650 & (NP{\_}001054;\ldots ;P02787) & disabled & \\
4.779e-01 & 2.205e-04 & KsVqEYANCHLAR & 1447.670 & (NP{\_}001054;\ldots ;P02787) & disabled & \\
\vdots & \vdots & \vdots & \vdots & \vdots & \vdots & \vdots  \\
7.524e-01 & 3.470e-04 & R$\ell$MNAsMVWAQAAR & 1448.720 & (NP{\_}000337;\ldots ;P48436) & disabled & \{($\ell$;2;108;Campomelic dysplasia (CMD1) [MIM:114290])\\ 
& & & & & & (s;6;112;Campomelic dysplasia (CMD1) [MIM:114290])
\} \\ \botrule
%\vdots & \vdots & \vdots & \vdots & \vdots & \vdots & \vdots \\ \botrule
\end{tabular} \vspace{-0.06in}
}
{RAId{\_}DbS by default perform searches considering the parent ion to have charge +1, +2, +3.
 The search results are then pooled together to form a single result ranked by $E$-values.
 This is why for the same spectrum RAId{\_}DbS may report peptide hits with very different masses.
 In part (a) ((b)), searches were done with annotated SAPs and PTMs turned off (on). 
The lowercase letters in the peptide indicate SAPs. A novel SAP, if present and enabled in the searches,
 will be specified in the column headed by Novel SAP. Note that in the disease related annotation, 
 there are four fields 
 separated by three semicolons. The first field, a lower case amino acid letter, 
 indicates the SAP; second field, an integer, indexes the SAP position in the peptide;  
 the third field, an integer, indexes the SAP position in the protein; the fourth field shows 
 the annotated disease association.  \vspace{-0.15in}
}
\end{table*}

\subsubsection*{RAId{\_}DbS Augmentation}
 Taking into account the finite sample effect and skewness, 
 the asymptotic score statistics ($P$-values) of RAId\_DbS~\citep{RAId_DbS} is {\it derived theoretically}. 
 The final $E$-value for each peptide hit, however, is obtained by multiplying the peptide's $P$-value
 by the number of peptides of its category. RAId\_DbS then ranks peptide hits according to $E$-values,
 not $P$-values. This avoids overstating the significance of a hit from a larger
 effective database and is particularly helpful in reducing false positives when we allow SAPs 
 and PTMs in the searches (see the supplementary information for details).

 In addition to performing database searches that consider annotated SAPs and PTMs, we also allow users
 to include for consideration one novel SAP per peptide. 
%We have augmented RAId{\_}DbS~\citep{RAId_DbS} 
% to perform database searches allowing annotated SAPs and PTMs.
% To accommodate possible new SAPs, we also allow users
% to elect the consideration of one novel SAP per peptide. 
%That is, RAId{\_}DbS will expand the search space on-the-fly to allow for one novel SAP,
% in addition to the annotated SAPs and PTMs, while perfroming the database searches,
% albeit at the expense of a longer run time. 
This feature, helpful for scrutinizing 
 otherwise unidentifiable spectra, will increase the effective peptide database size.
 However, it does not cause harm since the $E$-values associated with  
 peptide hits containing novel SAPs are obtained by multiplying their $P$-values
 with a much larger number than that for peptide hits without novel 
 SAPs (see supplementary information for details). 
 
%Thus the reported $E$-value for a peptide with novel SAPs is done by multiplying its $P$-value
 %by a large database size that is calculated via adding the number of qualified peptides with novel
 % SAPs and the number of qualified peptides without novel SAPs. Similarly, for a peptide without novel SAPs,
 %its reported $E$-value is calculated by multiplyiing its $P$-value only by the number of 
 %qualified peptides without SAP. 

%Another feature is the flexibility to  search in a user-specific database, where a user
% may input her/his own expert knowledge regarding SAPs and PTMs for certain proteins and 
% even put in new sequences. This flexibility is achieved by a simple database entry organization. 

\begin{figure}[b!]
\vspace{-0.12in}
%\begin{minipage}{0.6\columnwidth}
\textcolor{blue}{
\sf \scriptsize MLLATLLLLLLGGALAHPDRIIFPNHACEDPPAVLLEVQGTLQRPLVR$\la$\textcolor{red}{\bf \{W00\}}$\ra$D
SRTSPAN$\la$\textcolor{black}{\bf (N08,N09,N10,N11,N12)}$\ra$CTWLILGSKEQTVTIRFQKLHLACGSERL 
TLRSPLQPLISLCEAPPSPLQLPGGN$\la$\textcolor{black}{\bf(N08,N09,N10,N11,N12)}$\ra$VTITYSYAGA
RAPMGQGFLLSYSQDWLM$\la$\textcolor{red}{\bf \{V00\}}$\ra$CLQEEFQCLNHRCVSAVQR............[
%CDGVDAC
%GDGSDEAGCSSDPFPGLTPRPVPSLPCN$\la$\textcolor{blue}{\bf(N08,N09,N10,N11,N12)}$\ra$VTLEDFY
%GVFSSPGYTHLASVSHPQSCHWLLDPHDGRR$\la$\textcolor{red}{\bf\{W00\}}$\ra$LAVRF
%TALDLGFGDAVHVYDGPGPPE
%SSRLLRSLTHFSNGKAVTVETLSGQAVVSYHTVAWSNGRGFN$\la$(N08,N09,N10,N11,N12)$\ra$ATYHVRGYCLP
%WDRPCGLGSGLGAGEGLGERCYSEAQRCDGSWDCADGTDEEDCPGCPPGHFPCGA$\la$\{T00\}$\ra$AGTSGAT
%ACYLPADRCNYQTFCADGADERRCRHCQPGNFRCRDEKCVYETWVCDGQPDCADGSDEWDCS$\la$\{C00\}$\ra$
%YVLPRKVITAAVIGSLVCGLLLVIALGCTCKLYAIRTQEYSIFAPLSRMEAEIVQQQAPPSYGQLIAQGAIPPVEDF
%PTENPNDNSVLGNLRSLLQILRQDMTPGGGPGARRRQRGRLMRRLVR$\la$\{H00\}$\ra$RLRRWGLLPRTNTPARA
%SEARSQVTPSAAPLEALDGGTGPAREGGAVGGQDGEQAPPLPIKAPLPSASTSPAPTTVPEAPGPLPSLPLE
%PSLLSGVVQALRGRLLPSLGPPGPTRSPPGPHT$\la$\{R00\}$\ra$AVLALEDEDDVLLVPLAEPGVWVAEAEDEPLLT
} \vspace{-0.1in}
%\end{minipage}
\caption[]
{Protein sequence (NP{\_}054764) used as an example to demonstrate our database structure. 
A ``[" character is always inserted after the last amino acid of each protein to serve 
 as a separator. 
 Annotated SAPs and PTMs associated with an amino acid are included in a pair of angular brackets following
 that amino acid. SAPs are further enclosed by a pair of curly brackets while PTMs are 
 further enclosed by a pair of round brackets. Amino acid followed by two zeros indicates an annotated SAP. 
  Every annotated PTM has a two-digit positive integer 
 that is used to distinguish different modifications. 
 \vspace{-0.1in}
} \label{db_example}
\end{figure}
 
\vspace{-0.08in} 
\subsection*{Example}
Using a tandem mass (MS$^2$) spectrum taken from
 the profile dataset described earlier~\citep{E_calib}, 
 we illustrate in Table~1 two search results in the human protein database with the
 annotated SAPs and PTMs turned off (a) and on (b) respectively. 
  In case (a), the best hit is a false positive with $E$-value about $0.11$ implying
 that one probably ends up declaring no significant peptide hit for this spectrum.
 In case (b), however, the best hit is a true positive (a peptide from human {\it transferin}
 with an annotated SAP) with $E$-value about $4.0\times 10^{-7}$. This example shows that if properly used,
 allowing SAPs/PTMs may increase peptide identification rate. That is, it may be fruitful to turn on the 
 SAPs/PTMs when a regular search returns no significant hit. Blindly turn on SAPs/PTMs, however, may cause loss
 of sensitivity due to the increase of search space. In supplementary information, 
 using the $54$ training spectra of PEAKS, we compare RAId\_DbS's peptide identifications with and without SAPs/PTMs.
 The purpose is to study the degree of loss in senstivity when turning on the SAPs/PTMs. Although 
 for the data set tested there is no obvious loss in sensitivity (perhaps due to the 
 statistical accuracy of RAId\_DbS), we recommend running searches with SAPs/PTMs on only when 
  a regular search returns no significant hit.

\subsection*{Conclusion}
To enable speedy information retrieval and to enhance the protein coverage
  while analyzing MS$^2$ peptide spectra, we have augmented the capability of RAId{\_}DbS and 
 integrated with protein database additional information 
  such as SAPs, PTMs, and disease annotations. 
  Incorporation of known SAPs and PTMs during initial searches may enhance the peptide identification
 rate. Integration of disease knowledge and information may be crucial in many time-pressed clinical
  uses. 
% Although searching in a larger space than the regular cases, thanks to the accurate 
%spectrum-specific statistics provided by RAId{\_}DbS, one will not be
% seriously harmed by the increase of number of false positives.  

We are currently investigating the possibility of 
combining various isoforms of proteins into a single entry in addition to clustering almost identical proteins. 
 We are experimenting with keeping the longest form of the protein
 and marking at the beginning of the sequence possible deletions. 
Once achieved, this enhancement will further reduce
 {\it redundant} searches which should result in a shorter run time. 
 Another objective is to cover more organisms. Currently, we have finished database construction of $17$ 
 organisms, including {\it Homo sapiens, Drosophila melanogaster, Saccharomyces cerevisiae}  etc.
 (see supplementary information for details). We will provide more organismal databases
  on our web server once constructed.

\subsection*{Acknowledgement}
%We thank the administrative group of the NIH biowulf clusters,
%where all the computational tasks were carried out. 
This work was supported by
the Intramural Research Program of the National Library of Medicine
at National Institutes of Health/DHHS. Funding to pay the Open
Access publication charges for this article was provided by the NIH.

\newpage

\section*{Supplementary Information}

\subsection*{Information-preserved Protein Clustering and Database Construction} 
We extract $34,197$ human protein sequences with a total of $16,814,674$ amino acids
from the file (last updated 09/05/2006) 
 ftp://ftp.ncbi.nlm.nih.gov/genomes/H{\_}sapiens/protein/protein.gbk.gz. 
 Each protein sequence is accompanied by a list of annotated
  SAPs and PTMs.  
 Out of the $34,197$ proteins, we found $29,979$ unique proteins with a total of $15,324,913$ 
 amino acids. To avoid having multiple copies of identical or almost identical proteins 
 in the database, we first cluster the $34,197$ sequences by 
 running an all-against-all BLAST. %~\citep{Blast}. 
Two sequences with identical lengths
  and aligned gaplessly with less than $2\%$ mismatches are clustered together, and each sequence 
 is called a {\it qualified} hit of the other. 
 %For shorter sequences, when the $2\%$ mismatches
 %result in less than three mismatch positions, the condition is relaxed to allow up to three mismatches.   
  Any other sequence that satisfies this condition with a member of an existing 
 cluster is assigned to that existing cluster. All the annotations in the same
 cluster are then merged. We find it possible for every given
 cluster to choose a consensus sequence that will make all other 
 members its polymorphous forms. Hence, we only retain one protein sequence for
 each of the $29,272$  clusters. The total number of amino acids associated with these 
 $27,272$ consensus proteins is $15,001,326$.  
 %We end up with $29,272$ clusters. 

Although we only retain one sequence (the consensus sequence) per cluster, the information
 of other member sequences are still kept. For example, when a member sequence and the 
 consensus sequence disagree at two sites, the presence of the member sequence is documented 
 by introducing two {\it cluster-induced} SAPs at the two sites of the consensus sequence.
 The originally annotated SAPs and PTMs of the member sequence are also merged into those of
 the consensus sequence. Figure~\ref{cluster_example} illustrates how this is process is done
 iteratively. In our information file, each SAP or PTM is  documented with its origin. 
  SAPs arising from clustering are easily distinguished from annotated SAPs. For member sequences
 that are identical to the consensus sequence, the accession number of those member sequences
  are also recorded with their SAPs/PTMs annotations merged into the consensus sequence. 
 When a user selects not to have annotated SAPs, RAId\_DbS still allows for cluster-induced SAPs 
 resulting in an effective search of the original databases but with minimum redundancy.
 The strategy employed by RAId\_DbS to search for SAPs and PTMs will be briefly described  
  in the RAId\_DbS section below.

\begin{figure}[h!]
%\begin{minipage}{0.6\columnwidth}
\textcolor{black}{
\sf \scriptsize \hspace*{30pt}consensus seq. \hspace*{5pt} \ldots DPR\ldots\hspace{5.2pt}\ldots\hspace{5.2pt}\ldots L\textcolor{red}{\bf Q}RLV\textcolor{red}{\bf A}DN$\la$(N08)$\ra$GSE \ldots \\
\hspace*{36pt} member seq. \hspace*{5pt} \ldots DPR$\la$\{W00\}$\ra$\ldots L\textcolor{blue}{\bf K}RLV\textcolor{blue}{\bf V}DN$\la$(N11)$\ra$GSE \ldots \\
 updated consensus seq. \\  \hspace*{17pt} \ldots DPR$\la$\{W00\}$\ra$\ldots L\textcolor{red}{\bf Q}$\la$\{\textcolor{blue}{K00}\}$\ra$RLV\textcolor{red}{\bf A}$\la$\{\textcolor{blue}{V00}\}$\ra$DN$\la$(N08,N11)$\ra$GSE\ldots 
}
%\end{minipage}
\caption[]
{Information-preserved protein clustering example. 
 Once a consensus sequence  is selected, members of the clusters are
 merged into the consensus one-by-one. This figure illustrates how
  the information of a member sequence is merged into the consensus sequence.
 The difference in the primary sequences between a member and the consensus introduces
 {\it cluster-induced} SAPs. In this example, the residues Q and A (in red) in the consensus are 
 different from the residues K and V (in blue) in the member sequence. As a consequence, K becomes
 a cluster-induced SAP associated with Q and V becomes a cluster-induced SAP associated with A
 at these respective sites of the consensus. The annotated SAP, \{W00\}, associated with 
 residue R in the member sequence is merged into the consensus sequence, see the updated consensus
 sequence in the figure. Note that the annotated PTM, $\la$(N11)$\ra$, associated with N 
 in the member sequence is merged with a different annotated PTM, $\la$(N08)$\ra$, at the same
 site of the consensus sequence. As mentioned earlier, although the SAPs, PTMs are merged,
 each annotation's origin and disease associations are kept in the information file 
 allowing for faithful information retrieval at the final reporting stage of the RAId\_DbS 
 program. 
} \label{cluster_example}
\end{figure}

The consensus protein in a given cluster is then used as a query to BLAST against 
 the NCBI's nr database to retrieve its RefSeq accession number
 and its corresponding Swiss-Prot (http://ca.expasy.org/sprot/) accession number, 
 if it exists, from the best {\it qualified} hit.
  It is possible for a cluster to have more than one accession number. This happens
 when there is a tie in the qualified best hits and when a protein sequence
 in nr actually is documented with more than one accession number. 

To minimize inclusion of less confident annotations, we only keep the SAPs and PTMs that are
 consistently documented in more than one source. 
 For example, for proteins with Swiss-Prot accession number,
 we only keep the SAPs and PTMs that are annotated both by Swiss-Prot and GeneBank. 
 For proteins without Swiss-Prot accession numbers, the retentions of SAPs and
 PTMs are described below. The PTM annotations are kept 
 only if they are present in the gzipped document HPRD{\_}FLAT{\_}FILES{\_}090107.tar.gz of the
  Human Protein Reference Database: http://www.hprd.org/download. 
 The SAP annotations are kept only if they
 are in agreement with the master table, SNP{\_}mRNA{\_}pos.bcp.gz (last updated 01/10/2007), of dbSNP:
ftp://ftp.ncbi.nlm.nih.gov/snp/organisms/human{\_}9606/\\/database/organism{\_}data.

\vspace*{0.1in}
\subsection*{RAId{\_}DbS}
Taking into account the finite sample effect and skewness, 
 the form of asymptotic score statistics ($P$-values) of RAId\_DbS~\citep{RAId_DbS} is {\it derived theoretically}. 
Since the skewness varies {\it per spectrum}, the parameters for our theoretical distribution are 
 spectrum-specific. For each spectrum considered, our theoretical distribution (used to compute $P$-value)
  mostly agrees well with the score histogram accumulated. 
 The final $E$-value for each peptide hit, however, is obtained by multiplying the peptide's $P$-value
 by the number of peptides of its category. As a specific example, when Trypsin is used as the 
 digesting enzyme, RAId\_DbS allows for incorrect N-terminal cleavages. RAId\_DbS has internal counters, $C_c$
 and $C_{inc}$, counting respectively the number of scored peptides with correct and incorrect N-terminal cleavage. 
 In general, $C_{inc} \gg C_c$.  When calculating the $E$-value of a peptide with correct N-terminal cleavage, 
 RAId\_DbS multiplies the peptide's $P$-value by $C_c$. However, the $E$-value of a peptide with 
 incorrect N-terminal cleavage will be obtained by multiplying the peptide's $P$-value by $C_c+C_{inc}$~\citep{RAId_DbS}.   
 In line with the Bonferroni correction, our approach avoids overstating the significance of a hit from a larger
 effective database (the pool of peptides regardless of whether the N-terminal cleavage is correct) versus 
 a hit from a smaller effective database (the pool of peptides with correct N-terminal cleavage only). 
 
The same idea is used in the augmented RAId\_DbS. That is, different counters are set up to
 record the number of scored peptides in different categories. As a specific example, when novel SAPs are allowed,
 RAId\_DbS creates a new counter, $C_{novel\_sap}$, to record the number of scored peptides with a novel SAP. 
  This is in general a much larger number than other counters. When one calculates the $E$-value associated
 with a peptide hit that contains a novel SAP, one will multiply the peptide's $P$-value by the sum
 of a number of coutners with $C_{novel\_sap}$ included. However, in the same search, for a peptide
 without novel SAP, its $E$-value is obtained by multiplying the peptide's $P$-value by the sum of a number
 of counters {\it excluding} $C_{novel\_sap}$. The same approach is applied to PTMs and other annotations.

Below we briefly sketch how RAId\_DbS deals with the presence of annotated SAPs, PTMs as well as
 novel SAPs. In our database format, annotated SAPs and PTMs are inserted right after the 
 site of variation. When searching the database for peptides with parent ion mass 1500 Da, 
 RAId\_DbS sums the masses of amino acids within each possible peptide to see if the total mass is within 3 Da
 of $1500$ Da. At this stage, sites with variations will have, instead of a fixed mass, several possible
 masses depending on the number of SAPs/PTMs are annotated at these sites. Each peptide fragmnent covering
 some of those sites will therefore have several effective masses, each corresponding
 to a specific arrangement of SAPs/PTMs. If some of these masses happens to be
 within 3 Da of $1500$ Da, RAId\_DbS will score this peptide with corresponding annotated SAPs/PTMs that 
 give rise to the proper masses. If none of these masses are within the allowed molecular mass range,
 that peptide will not be scored. Note that this approach is computationally efficient in terms of 
  mass selection. For example, if a peptide contains a site with annotated SAPs/PTMs, one computes the 
 mass of this peptide by summing {\it once} the amino acid masses of other sites. It is then a simple matter to see
 whether the addition of this sum to the list of masses associated with the site with SAPs/PTMs 
  may fall in the desriable mass range. This approach is particularly powerful when there are more than one
 site with SAPs/PTMs in the peptide considered. The combinatorics associated with two sites with SAPs/PTMs
  only result in a longer list of possible masses to be added to the mass sum of unvaried sites.
 This should be constrasted with methods that incorporate SAPs via 
appending polymorphous peptides to the end of the primary sequence. In the latter approach,
  the program needs to do the mass sum multiple times, repeating the mass sum of unvaried sites, 
 and thus may significantly slow down the searches.

Despite RAId\_DbS's strategic advantage, introduction of SAPs/PTMs does increase the complexity of the 
 algorithm. Therefore, we limit per peptide the maximum number of annotated SAPs to be $2$ 
 and the maximum number of annotated PTMs to be $5$. To facilitate discovery, RAId\_DbS also 
 permits novel SAPs, but limited to one novel SAP per {\it not-yet-annotated} peptide, meaning
  peptides that do not contain any annotated SAPs/PTMs. 
 This is because the introduction of novel SAP largely expand the search space, and if one allows novel
 SAPs on peptides already documented with SAPs/PTMs, the search space expansion will be even larger
  and may render the search intractable. Currently, the novel SAP is expedited via a pre-computed list
 of amino acid mass difference. As an example, assume that one is searching for a peptide 
 with parent ion mass $1500$ Da, and a not-yet-annotated candidate peptide has mass $1477$ Da,
 $23$ Da smaller than the target mass. It happens that $23$ Da is also the mass difference between 
 Tryptophan and Tyrosine, and if the candidate peptide contains a Tyrosine, RAId\_DbS will replace that Tyrosine
 with a Tryptophan and score the new peptide. If the candidate peptide contains two Tyrosines, RAId\_DbS will replace 
 one Tyrosine at a time with a Tryptophan and score both the new peptides.    
 It is evident that the complexity grows fast if one were to allow for two novel SAPs per petpide.

\begin{figure}[!h]
\begin{center}
\includegraphics[width=0.92\columnwidth]{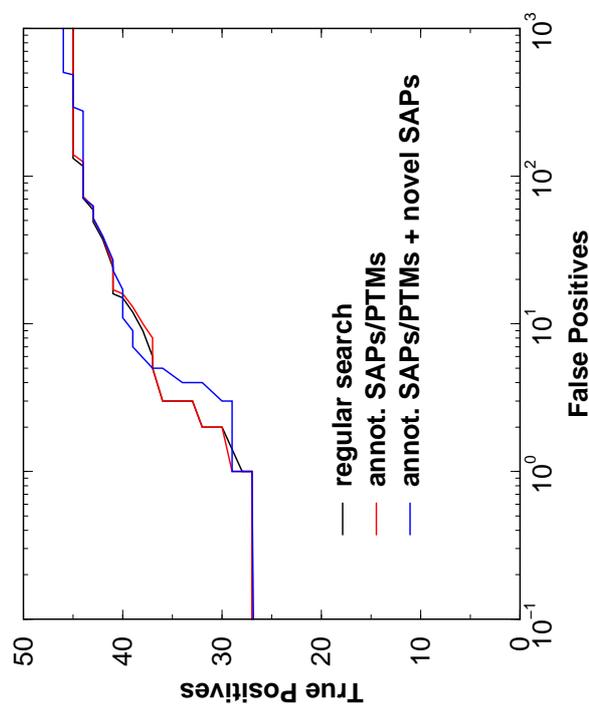}
%\vspace{-0.1in}
\end{center}
\caption[]{ROC curves for three different search strategies employed when running RAId\_DbS.  
} \label{fig.roc}
\end{figure}

It is commonly believed that when searching in a larger database, one is bound to loose sensitivity.
 This may be true if the $E$-value for every hit is obtained by multiplying the peptide's
 $P$-value by the same number regardless of the category that peptide belongs to. As we have explained 
 earlier, RAId\_DbS does not do that. It uses a method equivalent to Bonferroni correction. 
  We use $E$-values to rank peptide hits and each peptide's $E$-value is obtained by
 multiplying its $P$-value by the corresponding size of the effective database that the 
 peptide belongs to. Consequently,  peptide hits falling in a category that has a large effective database 
 size essentially need to have smaller $P$-values than those of peptide hits falling in a category that has a 
 small effective database size. In Figure~\ref{fig.roc}, we show the Receiver Operating Characteristic (ROC)
curves when analyzing the training $54$ spectra of PEAKS. In this data set, there are $17$ spectra from yeast,
 $23$ spectra from bovine, and $14$ spectra from horse. The true positive proteins are already provided by 
 PEAKS. We search the spectra generated by proteins of yeast, bovine, and horse respectively 
 in the databases fo yeast, bovine, and horse. Since the true positive proteins are already known, it is relatively
 easy to perform the ROC analysis using the search results from the $54$ spectra. 
 There are three ROC curves shown in Figure~\ref{fig.roc}, one for searches without SAPs/PTMs,
 one for searches allowing annotated SAPs/PTMs, and one for searches allowing both annotated SAPs/PTMs as well
 as novel SAPs.

\subsection*{Summary of Organismal Databases Constructed}
So far, we have finished constructing databses for $17$ organisms.  
We summarize these database in Table~\ref{tab.2} below. Note that the disease information
 is only available for the human database.  For human database, we have 
 $123,464$ SAPs and $81,984$ PTMs. Out of those SAPs and PTMs, $15,787$ of them have disease 
 associations.

\begin{table*}[th!]
\processtable{Summary of Augmented Organismal Databases Searchable by RAId{\_}DbS. \label{tab.2} \vspace{-0.1in}}
{
\begin{tabular}{llrrr}\toprule
Organism & DB\_name & SAPs included & PTMs included & DB\_size (byte)\\
{\it Homo sapiens} & hsa.seq &  123464 & 81984 &  16,292,193 \\
{\it Anopheles gambiae }& angam.seq & 350 & 50 & 6,042,277\\
{\it Arabidopsis thaliana} & artha.seq & 5207 & 11977 & 12,318,213 \\
{\it Bos taurus} & botau.seq & 3295 &  15810 & 11,188,490 \\
{\it Caenorhabditis elegans} & caele.seq & 1045 & 7756 & 10,050,609\\
{\it Canis familiaris}& cafam.seq & 2766 & 4196 & 18,458,474\\
{\it Danio rerio}& darer.seq & 7358 & 3841 & 14,477,794\\
{\it Drosophila melanogaster}& drmel.seq & 5611 & 9290 & 9,796,785\\
{\it Equus caballus}& eqcab.seq & 485 & 1045 & 9,404,150 \\
{\it Gallus gallus}& gagal.seq & 1109 & 6522 & 8,728,501 \\
{\it Macaca mulatta}& mamul.seq & 1370 & 1262 & 14,498,187\\
{\it Mus musculus}& mumus.seq & 27614 & 61684 & 14,363,491\\
{\it Oryza sativa}& orsat.seq & 1291 & 2182 & 10,679,924\\
{\it Pan troglodytes}& patro.seq & 5201 & 3734 & 20,227,873\\
{\it Plasmodium falciparum}& plfal.seq & 56 & 184 & 3,995,386\\
{\it Rattus norvegicus}& ranor.seq & 9297 & 33240 & 15,879,569\\
{\it Saccharomyces cerevisiae}& sacer.seq & 5507 & 13220 & 2,927,330\\
\botrule  
%\vdots & \vdots & \vdots & \vdots & \vdots & \vdots & \vdots \\ \botrule
\end{tabular} \vspace{-0.02in}
}
{ \vspace{-0.1in}
}
\end{table*}

\end{document}